\documentclass[
    ,final            
  ]
  {aipproc}

\layoutstyle{6x9}

\usepackage{bm}
\usepackage{graphicx,color,here}

\newcommand{\be}{\begin{equation}}
\newcommand{\ee}{\end{equation}}
\newcommand{\bea}{\begin{eqnarray}}
\newcommand{\eea}{\end{eqnarray}}

\begin{document}

\title{Exploring the transverse spin structure \\
of the nucleon}


\author{Umberto D'Alesio}{
  address={Dipartimento di Fisica, Universit\`a di Cagliari and INFN, Sezione di Cagliari, \\
  C.P 170, I-09042, Monserrato, Cagliari (Italy)}
}

\begin{abstract}
We discuss our present understanding of the transverse spin
structure of the nucleon and of related properties originating from
parton transverse motion. Starting from the transversity
distribution and the ways to access it, we then address the role
played by spin and transverse momentum dependent (TMD) distributions
in azimuthal and transverse single spin asymmetries. The latest
extractions of the Sivers, Collins and transversity functions are
also presented.
\end{abstract}

\maketitle


\subsection{Transversity and TMD distributions}
\label{intro}

\vspace*{-.1cm}

The study of the nucleon spin structure has made important progress
in the last years. Our understanding of the longitudinal degrees of
freedom, concerning both motion and spin content of partons inside
unpolarized and longitudinally polarized fast moving nucleons,
respectively encoded in the unpolarized quark distribution, $q(x)$,
and the helicity distribution, $\Delta q(x)=q_{+/+}-q_{-/+}$, is
quite accurate.

On the other hand the transversity distribution, $\Delta_T q(x) =
q_{\uparrow/\uparrow}-q_{\downarrow/\uparrow}$ (also denoted $h_1$),
despite its fundamental importance~\cite{Ralston:1979ys} and the
intense theoretical work of the last decade~\cite{Barone:2001sp},
has been accessed experimentally only very recently. The main
difficulty in measuring this function is that, being chiral-odd, it
decouples form inclusive deep inelastic scattering (DIS), since the
so-called handbag diagram cannot flip the chirality of quarks. The
only way to access this distribution is by coupling it to another
chiral-odd quantity. A further feature of $h_1$ is that it has no
gluonic counterpart, implying a non-singlet like QCD evolution. Very
recently an angular sum rule for the transversity, similar to that
for the helicity distribution, has been
proposed~\cite{Bakker:2004ib}. Despite some controversy it could be
extremely useful from a phenomenological point of view in
perturbative QCD.

In the usual collinear picture or when  all the intrinsic transverse
momenta ($k_\perp$) are integrated over, the three twist-two parton
distribution functions (PDF) $q,\,\Delta q$, and $ \Delta_T q$, give
a complete description of the structure of a nucleon. On the other
hand by taking into account the transverse motion a much richer
description can be obtained. At leading twist, the correlations
between spin and $k_\perp$ lead to five new distributions that
disappear when the hadronic tensor is integrated over $k_\perp$. For
a detailed classification and relations between different notations
for these functions see Refs.~\cite{Mulders:1995dh,
Bacchetta:2006tn, Bacchetta:2004jz, Anselmino:2005sh}. Here, by
restricting only to transverse spin
and transverse momentum correlations, we recall that: \\
$i$) two possible TMD distributions may contribute to the spin
asymmetry of transversely polarized quarks inside a transversely
polarized nucleon (the unintegrated transversity distribution
$h_1(x,k_\perp)$ and a new function, $h_{1T}^\perp(x,k_\perp)$;
$ii$) In a transversely polarized nucleon the azimuthal distribution of
an unpolarized quark can be asymmetric: this effect is encoded in
the chiral-even Sivers function~\cite{Sivers:1989cc,Sivers:1990fh},
denoted as $\Delta^N\!f_{q/p^\uparrow}$ (or $f_{1T}^\perp$); 
$iii)$ In an unpolarized nucleon a quark can be transversely polarized
orthogonally to the plane spanned by the quark and nucleon momenta:
this effect is encoded in the chiral-odd Boer-Mulders
function~\cite{Boer:1999mm}, denoted as $\Delta^N\!f_{q^\uparrow/p}$
(or $h_{1}^\perp$).

Analogously, in the fragmentation process into unpolarized hadrons,
besides the unpolarized fragmentation function (FF), $D_{h/q}$, a
new spin and TMD function appears: the chiral-odd Collins
function~\cite{Collins:1992kk}, $\Delta^N\! D_{h/q^\uparrow}$ or
$H_1^\perp$. This gives the azimuthal asymmetry in the fragmentation
of a transversely polarized quark into an unpolarized hadron.

The Sivers (Boer-Mulders) function is naively T-odd under time
reversal and this would imply its vanishing. On the other hand a
direct calculation~\cite{Brodsky:2002cx} in a quark-spectator model
showed that gluon exchange between the struck quark and the target
remnants allows a nonzero Sivers mechanism in DIS at leading twist.
As shown by Collins in Ref.~\cite{Collins:2002kn}, the gauge link
entering the operator definition of PDFs is crucial. Under time
reversal a future-pointing Wilson line, as probed in SIDIS, changes
into a past-pointing Wilson line, as probed in Drell-Yan (DY),
implying: 
$\Delta^N\!f_{q/p^\uparrow}|_{\rm DIS}\!=\!-\Delta^N\!f_{q/p^\uparrow}|_{\rm DY}$ (i.e.~a {\em modified
universality}). Concerning the Collins FF, not forbidden by time
reversal, in Ref.~\cite{Collins:2004nx} it has been shown that the
standard universality is preserved. This feature has been crucial in
the first extraction of transversity~\cite{Anselmino:2007fs}.

For completeness we mention another class of functions relevant in
our understanding of the nucleon structure, the generalized parton
distributions: for their general properties see
i.e.~Ref.~\cite{Diehl:2003ny}, while their connections to TMDs can
be found in Ref.~\cite{Meissner:2007rx}.

We discuss now how the study of transverse double spin asymmetries
(DSA) in the collinear picture and of transverse single spin
asymmetries (SSA) in a TMD approach could provide a powerful tool to
learn on the transverse spin structure of nucleons.

\subsection{Transverse double spin asymmetries} \label{dsa}

As already pointed out, to access $h_1$ in a physical observable we
need another chiral-odd partner. The most direct way, as originally
proposed in Ref.~\cite{Ralston:1979ys}, is via the transverse double
spin asymmetry in Drell-Yan processes, $p^\uparrow p^\uparrow
\rightarrow\ell^+\ell^- X$:
 \be
 \label{attdy}
 A_{TT} \equiv \frac{d\sigma^{\uparrow\uparrow} -
d\sigma^{\uparrow\downarrow}} {d\sigma^{\uparrow\uparrow} +
d\sigma^{\uparrow\downarrow}} = \hat a_{_{TT}} \> \frac{\sum_q e_q^2
\left[ h_1^q(x_1, M^2) \, h_1^{\bar q}(x_2, M^2)
 +  (1\leftrightarrow 2) 
 \right]}
{\sum_q e_q^2 \left[ q(x_1, M^2) \, \bar q(x_2, M^2)
 + (1 \leftrightarrow 2) 
 \right]}\, ,
 \ee
where the last expression refers to leading order (LO) accuracy, in
{\em collinear} pQCD, and where $M$ is the invariant mass of the
lepton pair and $\hat a_{_{TT}}$ is the elementary DSA for the $q\bar
q\rightarrow \ell^+\ell^-$ process. Here one measures the product of
two transversity distributions, one for a quark and one for an
anti-quark. At large energies, like those reachable at RHIC, NLO
corrections are small and LO expressions can be used. On the other
hand as shown in Ref.~\cite{Martin:1999mg}, $A_{TT}^{pp}$ in this
kinematical region is of the order of few percents: indeed, $h_1$
for antiquarks is expected to be small and at such small $x$ its
non-singlet nature implies a strong suppression compared to
$q(x,M^2)$ entering the denominator of $A_{TT}$.

A better way to access $h_1$ in DY processes is the transverse
double spin asymmetry in the collision of polarized protons and
antiprotons (implying a product of two {\em quark} transversity
distributions) in a kinematical region of intermediate $x$ values
(to have a sizeable $h_1^q$). These are the conditions of the
experiment proposed by the PAX collaboration~\cite{Barone:2005pu} at
GSI, where one expects $A_{TT}^{p\bar p}\simeq 20-40$\%. At such
moderate energies, NLO corrections, which are significant for the
unpolarized cross sections, almost cancel out in
$A_{TT}$~\cite{Shimizu:2005fp}. Moreover, to overcome the expected
low event rates, Ref.~\cite{Anselmino:2004ki} proposed to look at
the $J/\psi$ peak, where, gaining up to two orders of magnitude in
the yield, by reasonable assumptions one might still have a direct
access to $h_1$.

The inclusive production of photons or pions in $p^\uparrow
p^\uparrow$ collisions is another good channel to learn on $h_1$.
Here one expects larger rates compared to DY, but, due to the large
gluon contribution in the denominator, $A_{TT}$ might be very small.

Still in the collinear factorization, one can look for a chiral-odd
partner in the final hadron, like in $\ell
p^\uparrow\rightarrow\ell'\Lambda^\uparrow X$ or $p^\uparrow p
\rightarrow\Lambda^\uparrow X$. In SIDIS, for the $\Lambda$
polarization, we would have $ P_\Lambda \sim \sum_q h_1^q \otimes
\Delta_TD_{\Lambda/q}$. The advantage here is the self-analyzing
property of $\Lambda$ hyperons through their parity violating decay.
The price is twofold: 1) the appearance of a new unknown chiral-odd
function, $\Delta_TD_{\Lambda/q}$ (giving the probability that a
transversely polarized quark fragments into a transversely polarized
hadron); 2) the competing dominance of {\em up} quarks from the
incoming polarized hadron and of {\em strange} quarks in the spin
transfer to $\Lambda$ via the fragmentation process.

\subsection{Azimuthal and transverse single spin asymmetries}
\label{ssa}

As shown in Ref.~\cite{Kane:1978nd}, in collinear pQCD a transverse
SSA appears only as the imaginary part of interference terms between
helicity-flip and non-helicity-flip partonic scattering amplitudes.
Since at LO these are real and helicity is conserved for massless
partons, it was natural to expect (in $p^\uparrow p\rightarrow h X$,
for instance)
 \be
 A_N \equiv
\frac{d\sigma^\uparrow-d\sigma^\downarrow}{d\sigma^\uparrow+d\sigma^\downarrow}\simeq
\alpha_s m/\sqrt {\hat s}\,.
 \ee
Contrary to these expectations several experimental observations
show sizeable SSAs in the high-energy
regime~\cite{Adams:1991cs,Adams:1991ru,Adams:2003fx,Lee:2007zzh}.
The TMD approach to $p^\uparrow p\rightarrow \pi X$ was indeed
introduced to overcome this problem and a rich
phenomenology~\cite{Anselmino:1994tv,Anselmino:1998yz,D'Alesio:2004up,Anselmino:2005sh}
has been successfully developed. An alternative approach to these
SSAs, extending the QCD collinear factorization theorems to
higher-twist contributions, has been
formulated~\cite{Qiu:1998ia,Kouvaris:2006zy}. For an up-to-date overview
on SSAs see i.e.~Ref.~\cite{D'Alesio:2007jt}.

Before focusing on the phenomenology of TMD distributions, we
mention another tool to access the transversity distribution via
SSAs, but still in a collinear factorization picture.
Refs.~\cite{Jaffe:1997hf,Radici:2001na} proposed to consider
two-pion production in transversely polarized DIS: $\ell
p^\uparrow\rightarrow\ell\, \pi\, \pi\, X$. This allows $h_1$ to be
coupled to a new chiral-odd {\em interference} (or di-hadron) FF,
$\delta q_I$. This SSA has been observed by
HERMES~\cite{vanderNat:2005vt} and more information on $\delta q_I$
is going to be gathered by Belle from the study of
$e^+e^-\rightarrow h\, h\, X$.

The TMD approach applies naturally to leading-twist asymmetries,
like SSAs in Drell-Yan processes and SIDIS, or azimuthal asymmetries
in $e^+e^-\to h_1 h_2 X$, at low transverse momentum. For such
processes TMD factorization has been proved
\cite{Ji:2004xq,Ji:2004wu}.

Different groups have performed several phenomenological
analyses~\cite{Vogelsang:2005cs,Anselmino:2005ea,Anselmino:2005an,
Anselmino:2007fs,Anselmino:2008sga,Anselmino:2008sj,Efremov:2004tp,
Collins:2005ie,Collins:2005rq,Efremov:2006qm,Arnold:2008ap} and
their results are quite similar and consistent with each others. In
the following we will skip unessential details, like assumptions in
fitting procedures and experimental cuts, which can be found in the
above referred papers.

\subsubsection{Drell-Yan processes: $p\,p\rightarrow \ell^+\,\ell^-\, X$}

This process is a clean tool to learn on TMD distributions. Data on
unpolarized cross sections ($d\sigma/d\cos\theta d\phi_\ell$) show a
$\cos 2\phi$ dependence, that, if puzzling in LO and NLO collinear
QCD,  might be explained in terms of the convolution of two
Boer-Mulders functions~\cite{Boer:1999mm}. $\phi$ is the azimuthal
angle of the photon momentum in the Collins Soper lepton c.m.~frame
w.r.t.~the perpendicular lepton direction (identified by
$\phi_\ell$).

For transverse SSAs in DY, two mechanisms could play a role: the
Sivers effect and the Boer-Mulders effect (coupled to $\Delta_T q$).
Again in a schematic way we have
 \be
A_N \simeq \sum_q \big[\Delta^N\! f_{q/p^\uparrow} \otimes f_{\bar
q/p}\, \sin(\phi-\phi_S)  + \frac{\sin^2\theta}{(1+\cos^2\theta)}\,
 {\Delta_T q }\otimes \Delta^N\! f_{\bar q^\uparrow\!/p}\,
 \sin(\phi+\phi_S)\big]\,.
 \ee
This means that by proper integration over the lepton-pair angular
variables\footnote{Notice that the combination $(\phi-\phi_S)$,
related to the Sivers effect, does not depend on $\phi_\ell$.}
($\phi_\ell$) one can have a direct access to the Sivers
function~\cite{Anselmino:2002pd}, crucial to test the predicted
modified universality. At the same time this SSA represents another
tool to extract the transversity distribution. Experimental measurements
are planned (i.e.~RHIC, COMPASS).

\subsubsection{Azimuthal correlations in $e^+\,e^-\rightarrow h_1\,
h_2\, X$}

This process, even without any polarization, is the clearest way to
access the $k_\perp$-polarization correlation in the fragmentation
mechanism. The spins of the $q\bar q$ pair produced in the lepton
annihilation are strongly correlated and in the fragmentation into
two nearly back-to-back hadrons via the Collins mechanism a clear
azimuthal asymmetry might arise~\cite{Boer:1997mf}. This effect has
been observed by Belle at KEK~\cite{Seidl:2005bf,Seidl:2008xc}. In
particular, two experimental methods have been adopted: 1) by
reconstructing the 2-jet trust axis to look for a
$\cos(\phi_1+\phi_2)$ modulation ($\phi_{1,2}$ are the azimuthal
angles of the two hadrons w.r.t.~the plane spanned by the lepton
momenta and the thrust axis); 2) by observing the azimuthal
dependence of one hadron w.r.t.~the plane spanned by the other
hadron and the incoming beam, a $\cos (2\phi_0)$
dependence arises. In both cases one accesses the product of two Collins
functions. By suitable integrations, see
Ref.~\cite{Anselmino:2007fs}, one gets
 \bea
\label{ady}
 d\sigma &\simeq & \sum_q\big[ (1+\cos^2\theta)\,D_{h_1/q}(z_1)\,D_{h_2/\bar q}(z_2) \nonumber\\
 &+& \sin^2\theta \,\Delta ^N\! D _{h_1/q^\uparrow}(z_1)\,
 \Delta ^N\! D _{h_2/\bar q^\uparrow}(z_2)
\times \cos(\phi_1 + \phi_2)\big] \;\;\,\;[ {\it or\/}\;\times
\cos(2\phi_0)]\,.
 \eea
\begin{figure}[t!h]
\hspace*{2.cm}
\includegraphics[width=0.3\textwidth,bb= 110 240 540 760,
angle=-90] {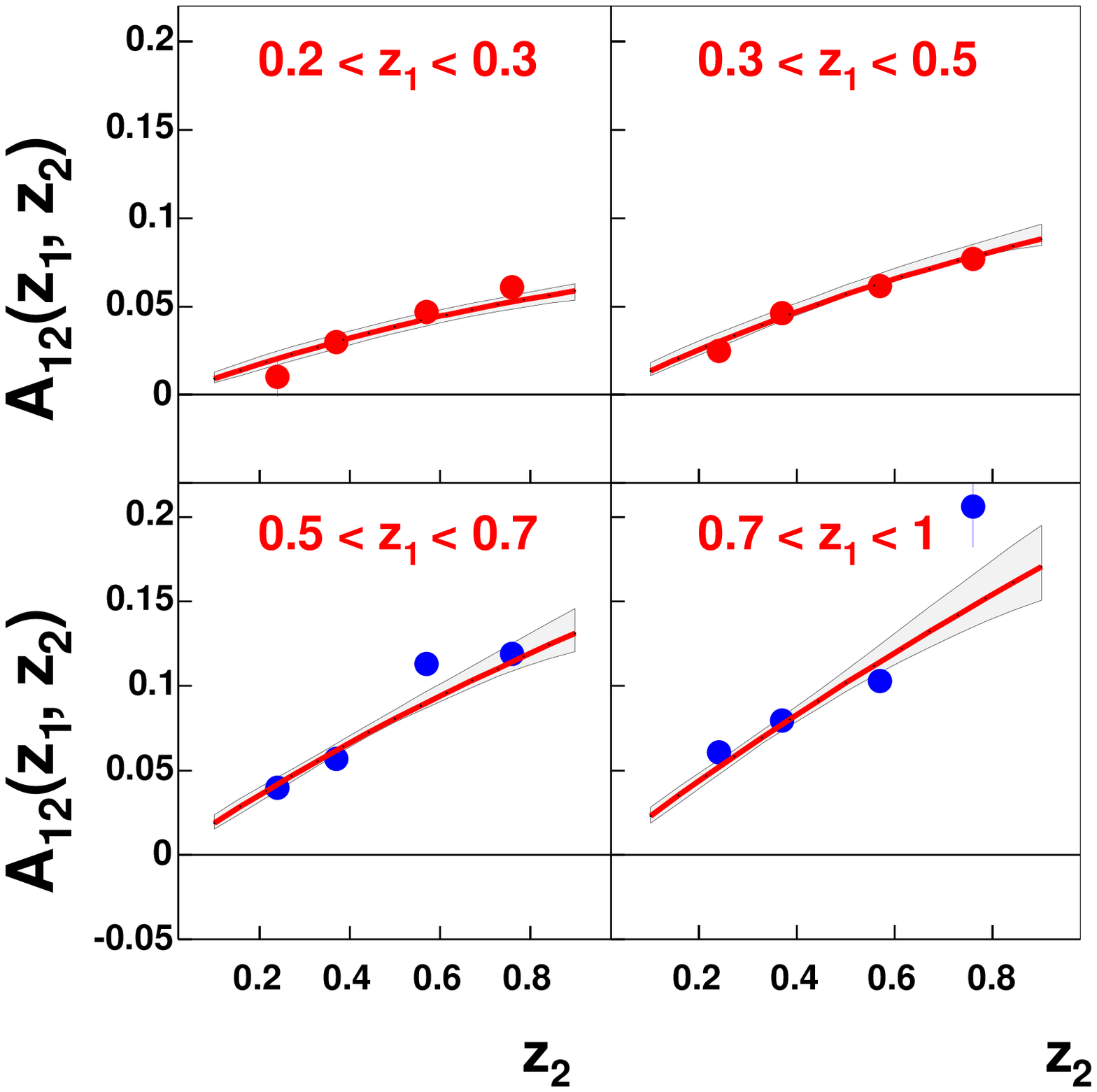} \hspace*{.4cm}
\includegraphics[width=0.3\textwidth,bb= 110 240 540 760,angle=-90]
{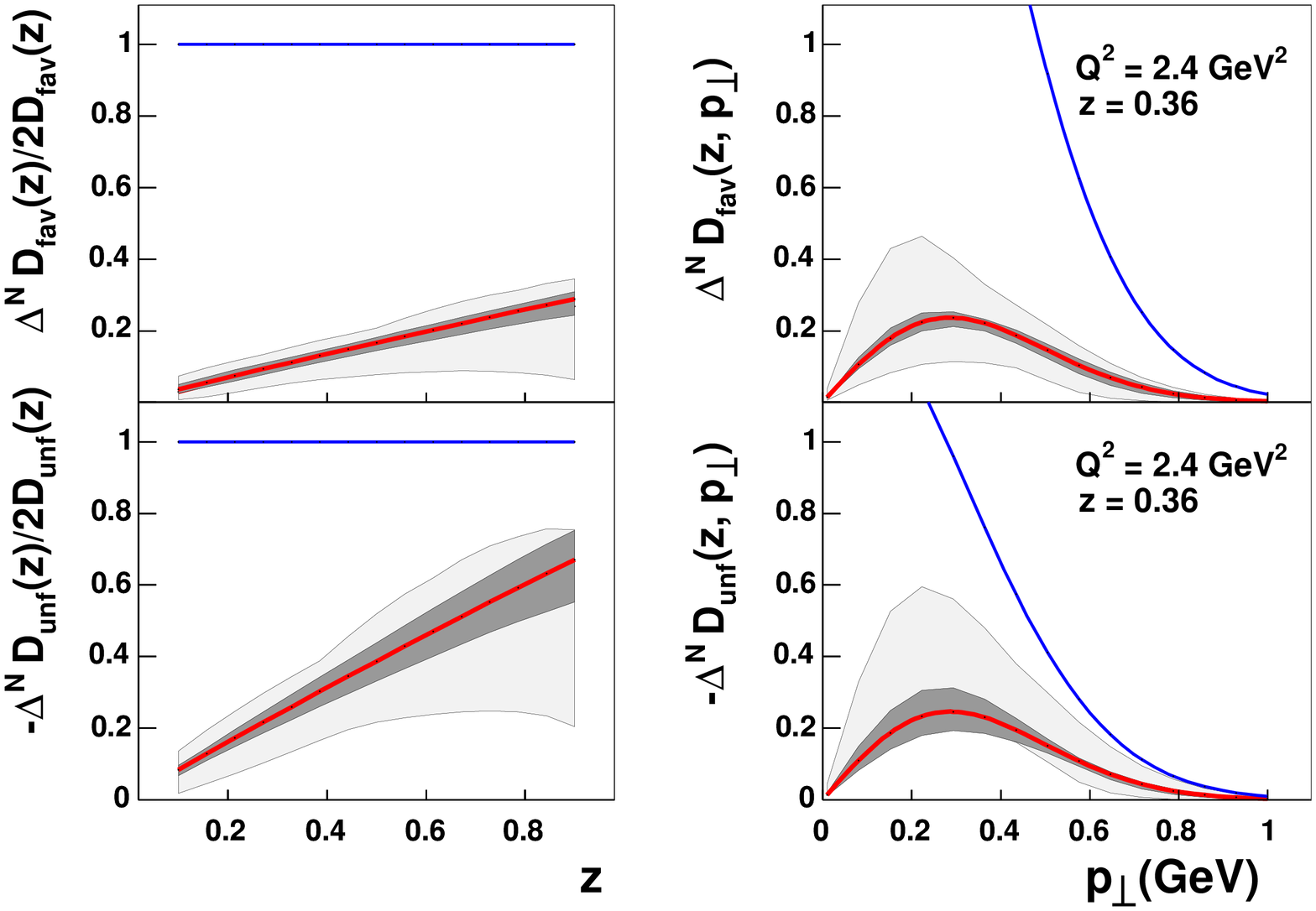}
 \caption{Left: Latest fit~\cite{Anselmino:2008sj} of the $\cos(\phi_1+\phi_2)$ modulation observed by Belle
 in $e^+e^-\rightarrow\pi\pi X$~\cite{Seidl:2008xc}. Right: Extracted favoured and unfavoured Collins
 functions. Shaded areas show the uncertainty on this extraction (wider bands correspond to the
 fit of Ref.~\cite{Anselmino:2007fs}).\label{coll}}
\end{figure}

In the study of Ref.~\cite{Anselmino:2007fs} the first data from
Belle~\cite{Seidl:2005bf} were used. New and more accurate data,
recently released, have allowed a better extraction of the Collins
function. In Fig.~\ref{coll} we show the preliminary updated
fit~\cite{Anselmino:2008sj} of the latest high statistics Belle
data~\cite{Seidl:2008xc} (left panel) and the corresponding favoured
(i.e.~$u\rightarrow \pi^+$) and unfavoured (i.e.~$d\rightarrow
\pi^+$) Collins functions (right panel). From these results we can
conclude that the Collins effect is sizeable, and favoured and
unfavoured Collins functions, comparable in magnitude, have opposite
sign. Notice that this information has been crucial in the global
analysis involving the Collins effect in SIDIS and leading to the
extraction of $h_1$.

\subsubsection{SIDIS processes: $\ell\, p\rightarrow \ell'\, h\, X$}
\label{sidis}

Also in this case the unpolarized cross section exhibits interesting
azimuthal dependences that could be explained in terms of TMD
distributions. In particular, the convolution of unpolarized
$k_\perp$-dependent PDFs and FFs gives rise to the so-called Cahn
effect~\cite{Cahn:1978se}. This has been indeed used for the
extraction of the $k_\perp$ behaviour of TMD
distributions~\cite{Anselmino:2005nn}. Moreover a $\cos 2\phi_h$
dependence could be ascribed to the convolution of the Boer-Mulders
function with the Collins FF~\cite{Boer:1997nt,Bacchetta:2006tn}.

Transverse SSAs in SIDIS have been the {\em gold} channel to access
for the {\em first time} the Sivers function and the transversity
distribution (this together with the Collins function by a combined
analysis of SIDIS and Belle $e^+e^-$ data). This has been possible
thanks to dedicated experimental programs by
HERMES~\cite{Airapetian:2004tw,Diefenthaler:2005gx,Diefenthaler:2007rj}
(with hydrogen target) and
COMPASS~\cite{Alexakhin:2005iw,Ageev:2006da,:2008dn} (with deuteron
target) collaborations. In particular, thanks to their latest high
statistics data~\cite{Diefenthaler:2007rj,:2008dn} we have been able
to improve our knowledge of the Sivers (including sea
quarks)~\cite{Anselmino:2008sga}, Collins and transversity
functions~\cite{Anselmino:2008sj}.

In a schematic form ({\em U}npolarized lepton, {\em T}ransversely
polarized target) we have
 \be
 \label{aut}
A_{UT} \simeq \sum_q\big[\Delta^N\! f_{q/p^\uparrow} \otimes D_{h/q}
\sin(\phi_h-\phi_S) + \frac{1-y}{1+(1-y)^2}\,\Delta_T q \otimes
\Delta^N\! D_{h/q^\uparrow} \sin(\phi_h+\phi_S)\big] + \cdots\,,
 \ee
where $\phi_h$ ($\phi_S$) is the azimuthal angle of the observed
hadron momentum (target polarization vector) in the photon-nucleon
c.m.~frame, measured from the leptonic plane~\cite{Bacchetta:2004jz}. An
extra term involving again the Collins function coupled to the
$h_{1T}^\perp$ distribution and providing a $\sin(3\phi_h-\phi_S)$
azimuthal modulation has been omitted.

The different azimuthal dependence in Eq.~\ref{aut} allows the
separation of Sivers and Collins effects. It is therefore common to
consider the following azimuthal moments
  \be
 \label{exp:sidis:aut} A_{UT}^{\sin(\phi_h \pm \phi_S)}
 = 2 \,\frac{\int d\phi_h d\phi_S
\sin(\phi_h \pm \phi_S) [d\sigma(\phi_S) -
d\sigma(\phi_S+\pi)]}{\int d\phi_h d\phi_S
  [d\sigma(\phi_S) + d\sigma(\phi_S+\pi)]}\,.
 \ee

\begin{figure}[t]
\hskip .2cm
\includegraphics[width=0.35\textwidth,bb= 10 140 540 660,angle=-90]
{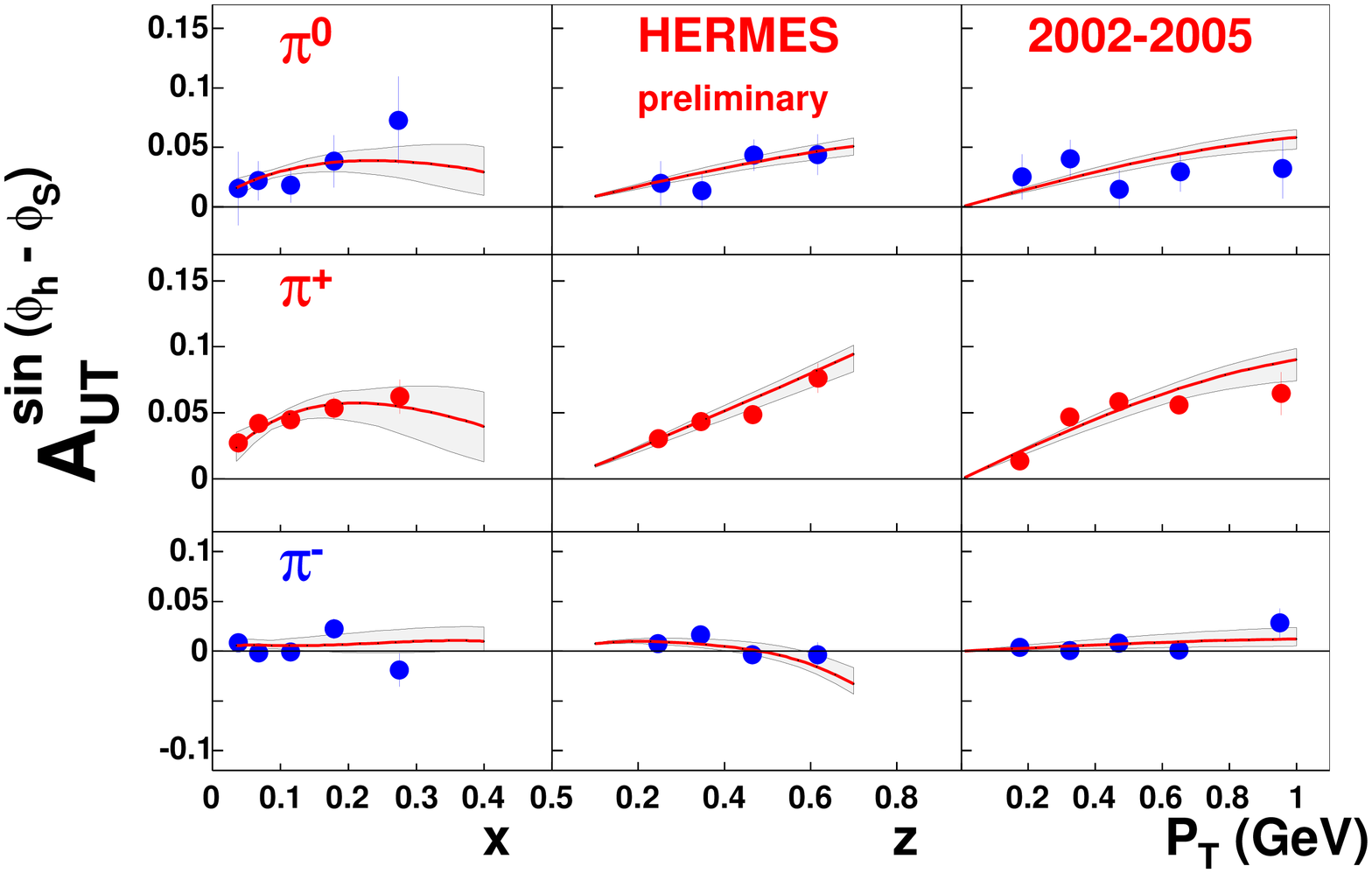} \hspace*{2cm}
\includegraphics[width=0.35\textwidth,bb= 10 140 540 660,angle=-90]
{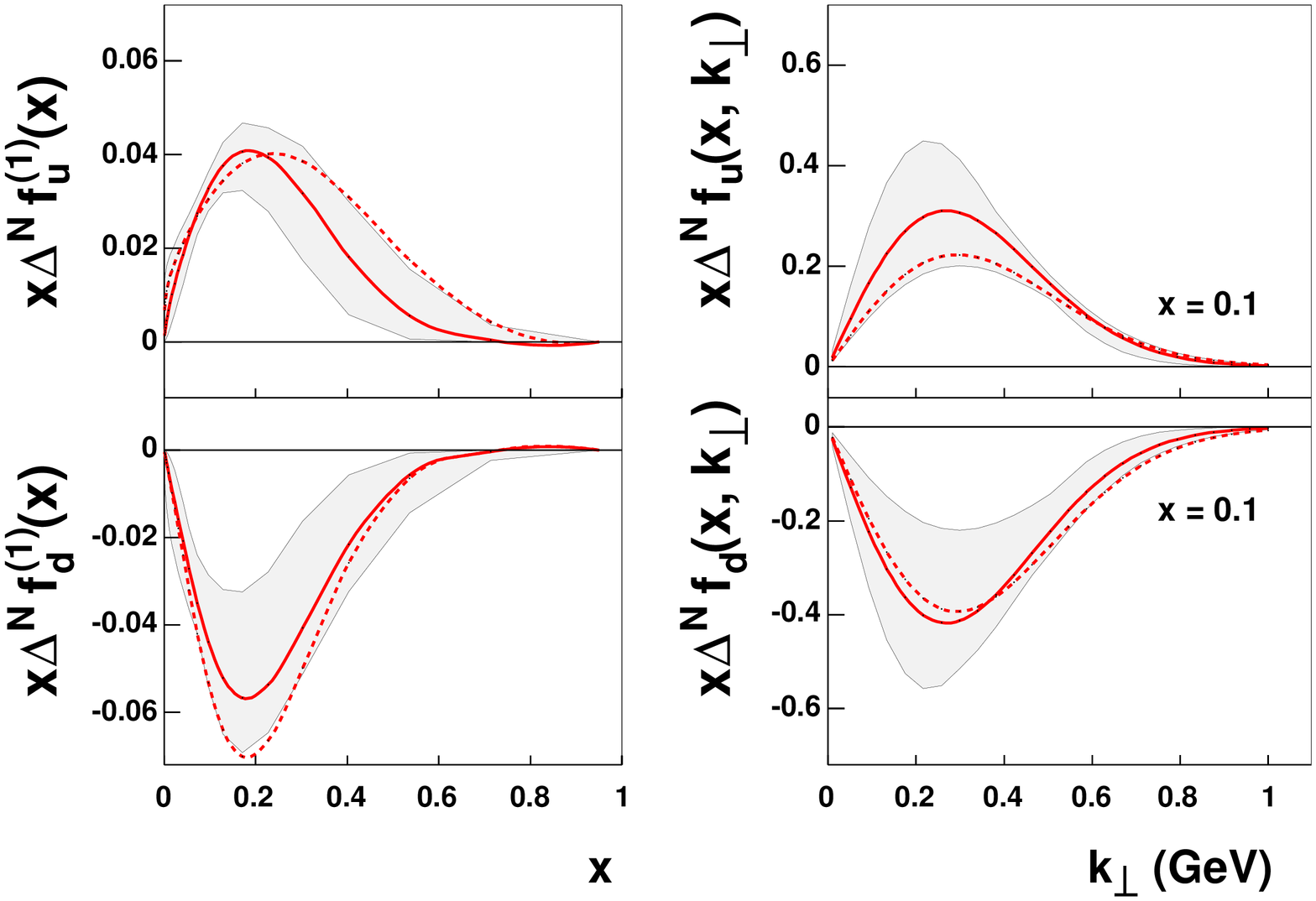}
 \caption{Left: Fit~\cite{Anselmino:2008sga} of HERMES $A_{UT}^{\sin(\phi_h-\phi_S)}$
 data~\cite{Diefenthaler:2007rj}. Right: Sivers function (and its first $k_\perp$-moment) for up and down quarks
 (solid curves) compared with results (dashed curves) from our previous analysis~\cite{Anselmino:2005ea}.
 Uncertainty bands are also shown. \label{aut-siv}}
\end{figure}

In Fig.~\ref{aut-siv} (left) we show the latest
analysis~\cite{Anselmino:2008sga} of  Sivers asymmetry data from
HERMES~\cite{Diefenthaler:2007rj} and (right) the Sivers function
for up and down quarks (solid lines) compared with the results from
our previous study (dashed lines)~\cite{Anselmino:2005ea}.

\begin{figure}
\includegraphics[width=0.35\textwidth,bb= 10 140 540 660,angle=-90]
{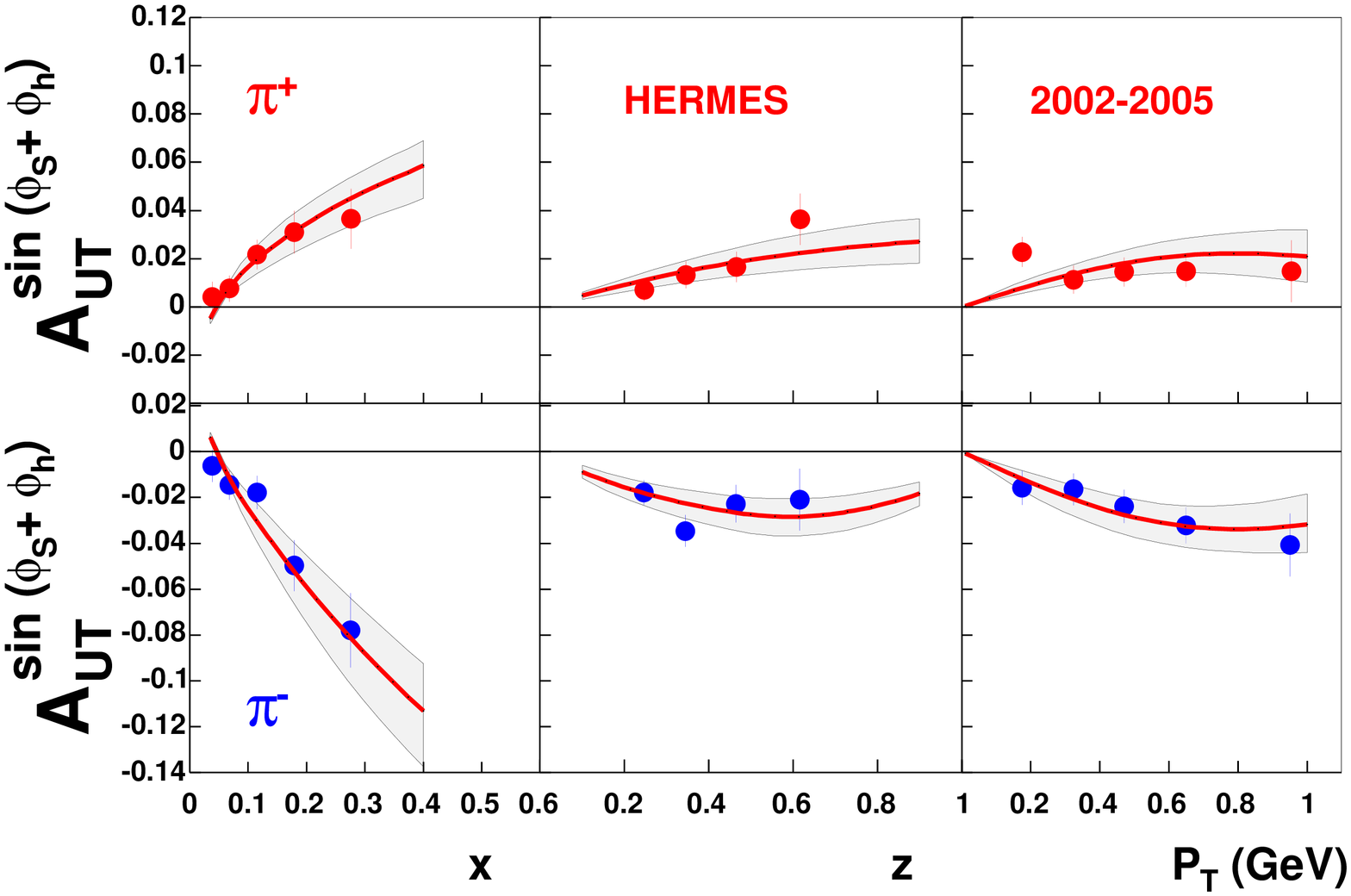} \hspace*{2cm}
\includegraphics[width=0.35\textwidth,bb= 10 140 540 660,angle=-90]
{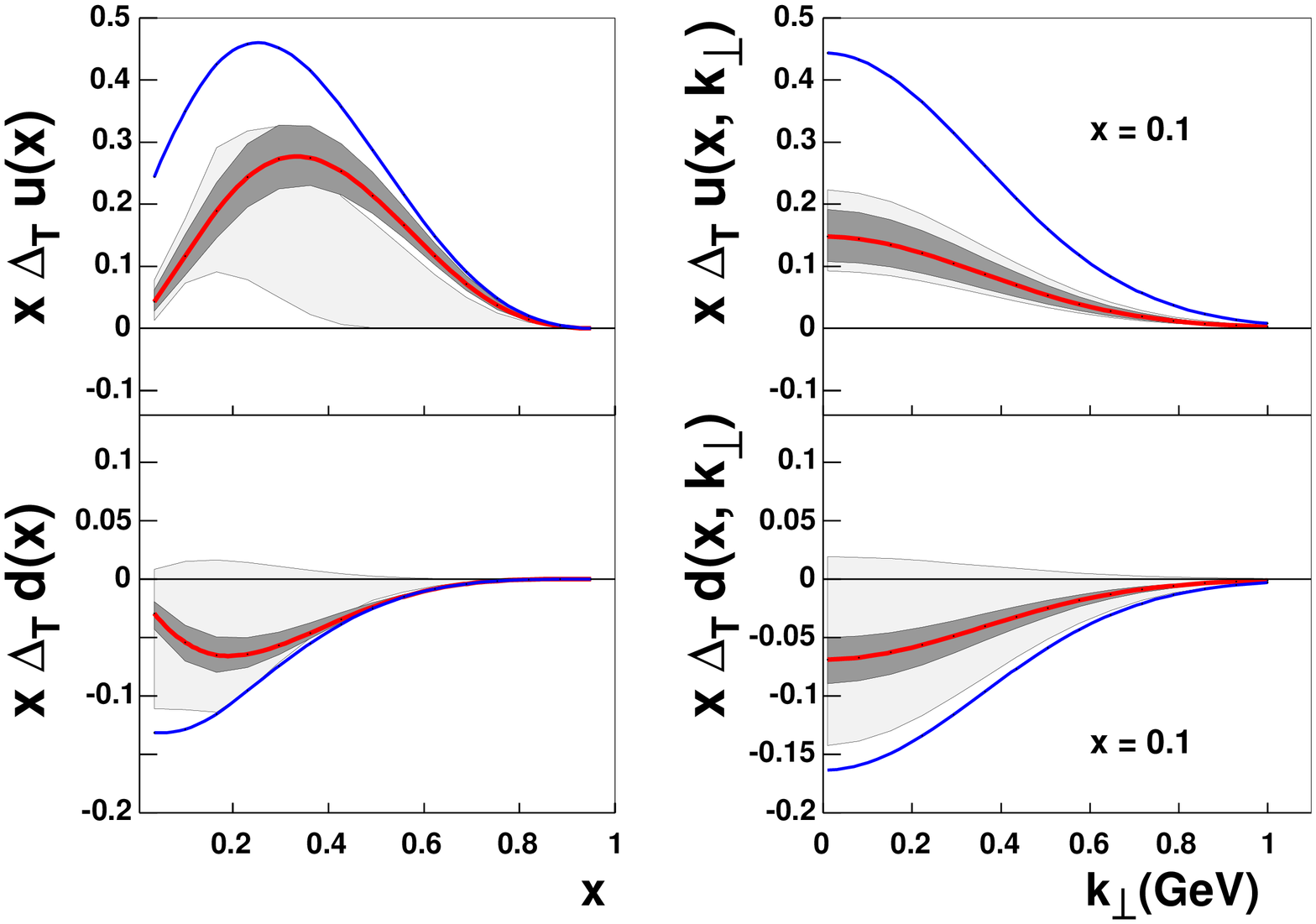} \caption{Left: updated
fit~\cite{Anselmino:2008sj} of HERMES $A_{UT}^{\sin(\phi_h+\phi_S)}$
data~\cite{Diefenthaler:2007rj}. Right: transversity distribution
for up and down quarks. For comparison we show the wider error bands
from our previous analysis~\cite{Anselmino:2007fs} and the Soffer
bound~\cite{Soffer:1994ww} (bold lines).\label{aut-col}}
\end{figure}

In Fig.~\ref{aut-col} (left) we show the updated
fit~\cite{Anselmino:2008sj} of Collins asymmetry data from
HERMES~\cite{Diefenthaler:2007rj} and (right) the extracted
transversity distribution for up and down quarks.

Notice that COMPASS SSA data on deuteron, even if compatible with
zero, complement with HERMES results concerning flavour separation
of TMDs.

From these analyses we see that: 1) The Sivers functions for valence
quarks are sizeable and opposite in sign for up and down quarks; 2)
Transversity is different from zero, smaller than its Soffer
bound~\cite{Soffer:1994ww}, with up quarks carrying a degree of
transverse polarization bigger (and opposite in sign) than down quarks.

\subsubsection{SSAs in $pp\rightarrow h\, X$}
Despite a clear experimental evidence and a rich phenomenology in
the TMD approach, this case might deserve further theoretical study.
Assuming a TMD factorization, both the Sivers~\cite{D'Alesio:2004up}
and the Collins~\cite{Yuan:2008tv} mechanism might explain the large
SSAs observed in $pp\rightarrow \pi\, X$. The extension of the
universality of the Collins function to $pp$ collisions is presented
in Ref.~\cite{Yuan:2007nd}, while, even though at a phenomenological
level, Ref.~\cite{Boglione:2007dm} shows the compatibility of the
Sivers effect in SIDIS and $pp\rightarrow\pi X$. For such a process
the two mechanisms are not separable and information from less
inclusive reactions, like $p^\uparrow p\rightarrow \gamma\; {\rm
jet}\; X$ ($\Delta^N\!f_{q/p^\uparrow}$)~\cite{Bacchetta:2007sz} or
$p^\uparrow p\rightarrow \pi\; {\rm jet}\; X$ ($h_1\otimes \Delta^N
\! D_{\pi/q^\uparrow}$)~\cite{Yuan:2007nd}, could help.

\vspace*{0.5cm}

The effective investigation of the nucleon transverse spin has
definitely started. Much progress has been made and further work is
needed both theoretically (a more solid picture of SSAs in the TMD
approach, $Q^2$-evolution of TMDs) and experimentally (DSAs and SSAs
in DY processes, data at large $x$ and for less inclusive
processes).

\vspace*{0.5cm}

I am deeply in debt with M.~Anselmino, M.~Boglione, A.~Kotzinian,
E.~Leader, S.~Melis, F.~Murgia and A.~Prokudin for their invaluable
collaboration.

\bibliographystyle{aipproc}   
\bibliography{spiresppnp}

\end{document}